\documentclass[%
reprint,
twocolumn,
superscriptaddress,
showpacs,preprintnumbers,
nofootinbib,
amsmath,amssymb,
aps,
prl,
]{revtex4}
\usepackage{hyperref}

\usepackage{graphicx}

\usepackage[dvips]{color}

\usepackage{ulem}

\def\al{\alpha} 
\def\be{\beta}

\def\ep{\epsilon}

\def\ka{\kappa}

\def\Om{\Omega}

\def\pa{\partial}

\newcommand{\Tfield}{\tau^\phi}
\newcommand{\Tfluid}{\tau^\text{f}}

\newcommand{\ben}{\begin{equation}}
\newcommand{\een}{\end{equation}}
\newcommand{\bea}{\begin{eqnarray}}
\newcommand{\eea}{\end{eqnarray}}
\newcommand{\ba}{\begin{array}}
\newcommand{\ea}{\end{array}}
\newcommand{\bit}{\begin{itemize}}
\newcommand{\eit}{\end{itemize}}

\newcommand{\mpl}{m_{\text{P}}}

\newcommand{\half}{\frac12}
\newcommand{\pdof}{g}  % Pressure degrees of freedom
\newcommand{\TN}{T_\text{N}} % Nucleation temperature
\newcommand{\tN}{t_\text{N}} % Nucleation temperature
\newcommand{\Tc}{T_\text{c}} % Critical temperature
\newcommand{\vw}{v_\text{w}} % Wall velocity
\newcommand{\fieldV}{\overline{U}_\phi} % Mean field gradient
\newcommand{\fluidV}{\overline{U}_\text{f}}  % Mean velocity
\newcommand{\Rb}{R} % Droplet radius
\newcommand{\Rbc}{R_*} % Mean droplet radius at collision
\newcommand{\Hc}{H_*} % Hubble rate at transitions
\newcommand{\Ts}{\tau_\text{s}} % Lifetime of acoustic oscillations
\newcommand{\etaS}{\eta_\text{s}} % Shear viscosity
\newcommand{\Tvisc}{\tau_{\eta}} % Lifetime due to Shear viscosity
\newcommand{\Nb}{N_\text{b}}

\newcommand{\tCohPar}{x_\text{c}}

\newcommand{\quadPar}{\gamma}
\newcommand{\cubPar}{\alpha}
\newcommand{\relgamma}{W}
\newcommand{\strengthPar}[1]{\alpha_{#1}}

\begin{document}

\newcommand{\Sussex}{\affiliation{
Department of Physics and Astronomy,
University of Sussex, Falmer, Brighton BN1 9QH,
U.K.}}

\newcommand{\HIPetc}{\affiliation{
Department of Physics and Helsinki Institute of Physics,
PL 64, % (Gustaf H\"{a}llstr\"{o}min katu 2),
FI-00014 University of Helsinki,
Finland
}}

\title{Gravitational waves from the sound of a first order phase transition}
\author{Mark Hindmarsh}
\email{m.b.hindmarsh@sussex.ac.uk}
\Sussex
\HIPetc
\author{Stephan J. Huber}
\email{s.huber@sussex.ac.uk}
\Sussex
\author{Kari Rummukainen}
\email{kari.rummukainen@helsinki.fi}
\HIPetc
\author{David J. Weir}
\email{david.weir@helsinki.fi}
\HIPetc

\date{January 30, 2014}
\begin{abstract}
We report on the first 3-dimensional numerical simulations of
first-order phase transitions in the early universe to include the
cosmic fluid as well as the scalar field order parameter.  We
calculate the gravitational wave (GW) spectrum resulting from the
nucleation, expansion and collision of bubbles of the low-temperature
phase, for phase transition strengths and bubble wall velocities
covering many cases of interest.  We find that the compression waves
in the fluid continue to be a source of GWs long after the bubbles
have merged, a new effect not taken properly into account in previous
modelling of the GW source.  For a wide range of models the main
source of the GWs produced by a phase transition is therefore the
sound the bubbles make.
\end{abstract}
\preprint{HIP-2013-07/TH}
\maketitle

In a hot Big Bang there were phase transitions in the early Universe~\cite{Kirzhnits:1972iw,Kirzhnits:1976ts}, which may well have been of first order; one major consequence of such a transition would be the generation of gravitational waves~\cite{Witten:1984rs,1986MNRAS.218..629H,Kosowsky:1991ua,Kosowsky:1992rz,Kosowsky:1992vn,Kamionkowski:1993fg}. The electroweak transition in the Standard Model is known to be a cross-over~\cite{Kajantie:1996mn,Laine:1998vn,Laine:2012jy}
but it may be first order in minimal extensions of the Standard Model~\cite{Carena:1996wj,Delepine:1996vn,Laine:1998qk,Grojean:2004xa,Huber:2000mg,Huber:2006wf}. It is therefore essential to properly characterise the expected power spectrum from first-order phase transitions.

First order phase transitions proceed by the nucleation, growth, and merger of bubbles of the low temperature phase~\cite{Steinhardt:1981ct,Witten:1984rs,KurkiSuonio:1984ba,Kajantie:1986hq,Enqvist:1991xw,Ignatius:1993qn,KurkiSuonio:1995vy,KurkiSuonio:1996rk,Espinosa:2010hh}.  The collision of the bubbles is a violent process, and both the scalar order parameter and the fluid of light particles generate gravitational waves.

Numerical studies have been carried out of the behaviour of bubbles in such a phase transition using spherically symmetric $(1+1)$-dimensional simulations~\cite{KurkiSuonio:1995vy,KurkiSuonio:1996rk}. 
The calculation of the gravitational wave spectrum has been refined in the intervening years, notably using 
the semi-analytic envelope approximation ~\cite{Kosowsky:1991ua,Kosowsky:1992vn,Kamionkowski:1993fg,Huber:2008hg,Caprini:2009fx} (but see Ref.~\cite{Caprini:2007xq} for an alternative approach).
Fully three-dimensional simulations of the scalar field only have been carried out~\cite{Child:2012qg}, qualitatively supporting the envelope approximation, and pointing out important gravitational wave production from the scalar field after the bubble merger.

In a hot phase transition, the fluid plays an important role, firstly as a brake on the scalar field, and secondly as a source of gravitational waves itself. The fluid has generally been assumed to be incompressible and turbulent~\cite{Kosowsky:2001xp,Gogoberidze:2007an,Caprini:2006jb,Caprini:2009yp}. An important question for the gravitational wave power spectrum is 
the validity of this modelling, which generally borrows from the Kolmogorov theory of non-relativistic driven incompressible turbulence.

In this Letter we report on the first fully three dimensional
simulation of bubble nucleation involving a coupled field-fluid
system.  We make use of these simulations to calculate the power
spectrum of gravitational radiation from a first-order phase
transition, for a range of transition strengths and bubble wall
velocities relevant for an electroweak transition in extensions of the
Standard Model.  We find that the compression waves in the fluid --
{sound waves} -- continue to be an important source of gravitational
waves for up to a Hubble time after the bubble merger has
completed. This boosts the signal by the ratio of the Hubble time to
the transition time, which can be orders of magnitude.

The system describing the matter in the early universe consists of a
relativistic fluid coupled to a scalar field, which acquires an
effective potential
\begin{equation}
V(\phi, T) = \frac{1}{2} \quadPar (T^2-T_0^2) \phi^2 - \frac{1}{3} \cubPar T \phi^3 + \frac{1}{4}\lambda\phi^4.
\end{equation}
The rest-frame pressure $p$ and energy density $\epsilon$ are
\begin{equation}
\epsilon = 3 a T^4 + V(\phi,T) - T\frac{\partial V}{\partial T}, \quad
p = a T^4 - V(\phi,T)
\end{equation}
with $a=(\pi^2/90)\pdof$, and $\pdof$ the effective number of
relativistic degrees of freedom contributing to the pressure at
temperature $T$. The stress-energy tensor for a scalar field $\phi$
and an ideal relativistic fluid $U^\mu$ is
\begin{equation}
\label{eq:tmunu}
T^{\mu\nu} = \partial^\mu \phi \partial^\nu \phi - {\textstyle \half} g^{\mu\nu} (\partial\phi)^2
+ \left[\epsilon + p \right] U^\mu U^\nu + g^{\mu\nu} p
\end{equation}
where the metric convention is $(-\,+\,+\,+)$. The scalar field
potential is included in the definition of $p$. We split $\pa_\mu
T^{\mu\nu}=0$ (nonuniquely) into field and fluid parts with a
dissipative term permitting transfer of energy between the scalar
field and the fluid $\delta^\nu = \eta U^\mu \partial_\mu \phi
\partial^\nu \phi$~\cite{Ignatius:1993qn,KurkiSuonio:1995vy}. This
simplified model can be improved, but is adequate for parametrising
the entropy production \cite{KurkiSuonio:1996rk}.

Given these expressions, the equations of motion can be derived. For the field we have
\begin{equation}
- \ddot{\phi} + \nabla^2 \phi - \frac{\partial V}{\partial \phi} = \eta \relgamma (\dot{\phi} + V^i \partial_i \phi)
\end{equation}
where $\relgamma$ is the relativistic $\gamma$-factor and $V^i$ is the
fluid 3-velocity, $U^i = \relgamma V^i$. For the fluid energy density
$E=\relgamma\epsilon$, contracting $\left[\partial_\mu
  T^{\mu\nu}\right]_\text{fluid}$ with $U_\nu$ yields
\begin{multline}
\dot{E} + \partial_i (E V^i) + p [\dot{\relgamma} + \partial_i (\relgamma V^i)]  - \frac{\partial V}{\partial \phi} \relgamma (\dot{\phi} + V^i \partial_i \phi) \\ = \eta \relgamma^2 (\dot{\phi} + V^i \partial_i \phi)^2.
\end{multline}
The equations of motion for the fluid momentum density $Z_i =
\relgamma(\epsilon + p)U_i$ read
\begin{equation}
\dot{Z}_i + \partial_j(Z_i V^j) + \partial_i p + \frac{\partial V}{\partial \phi} \partial_i \phi  = -\eta \relgamma (\dot{\phi} + V^j \partial_j \phi)\partial_i \phi.
\end{equation}

\begin{figure}[t]

\begin{centering}
\includegraphics[scale=0.1]{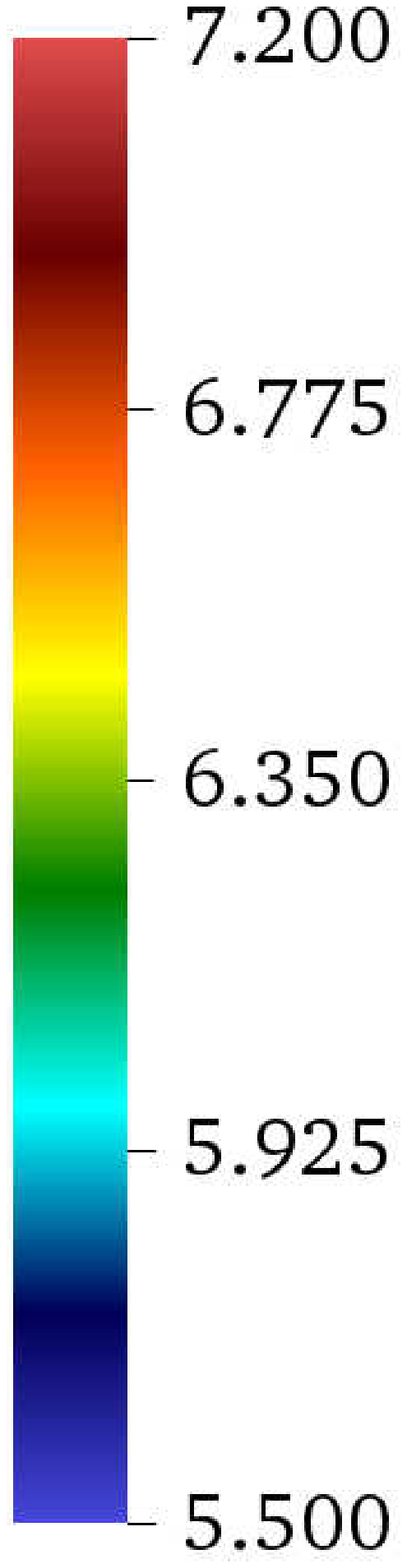}\hspace{5px}
\includegraphics[scale=0.1]{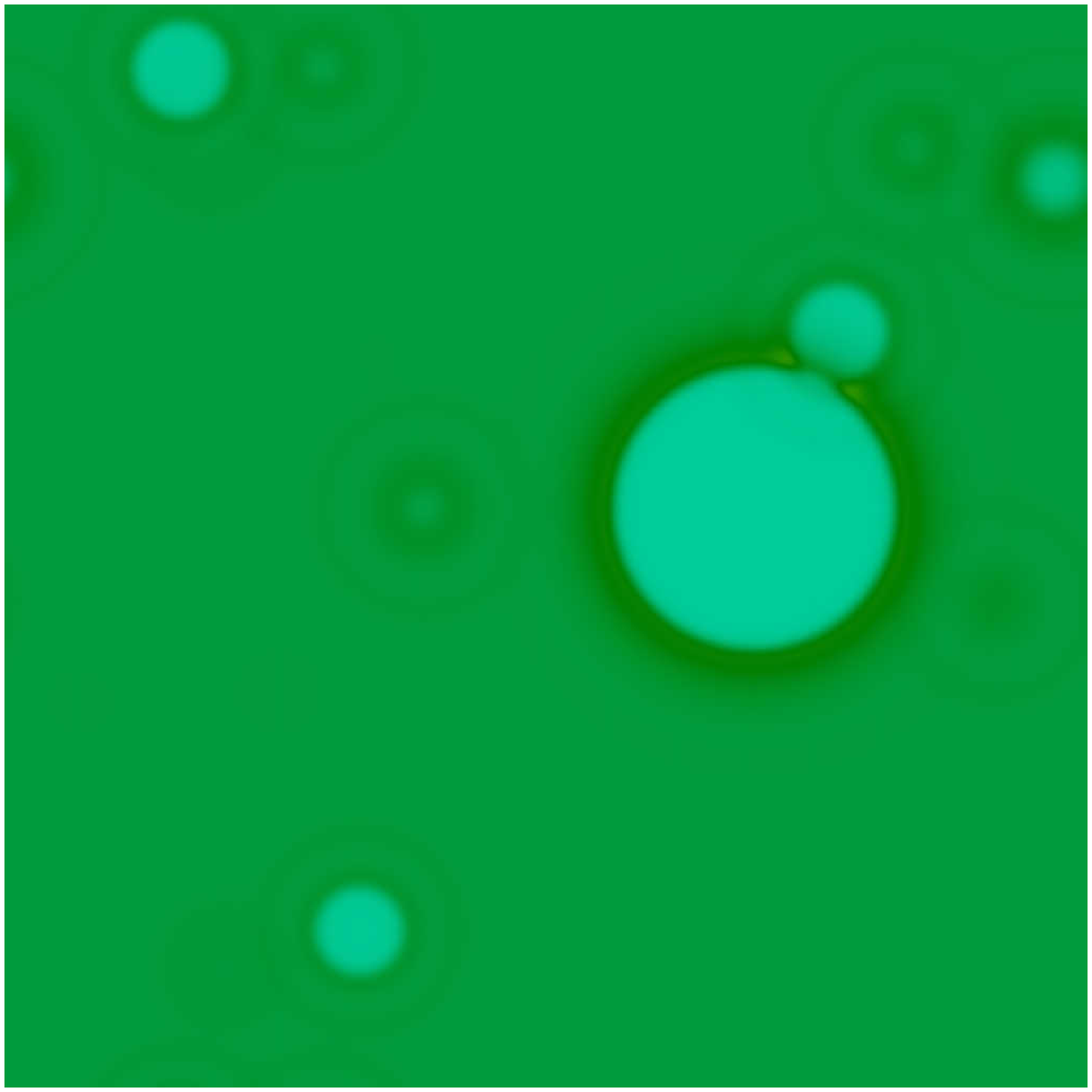}\hspace{5px}
\includegraphics[scale=0.1]{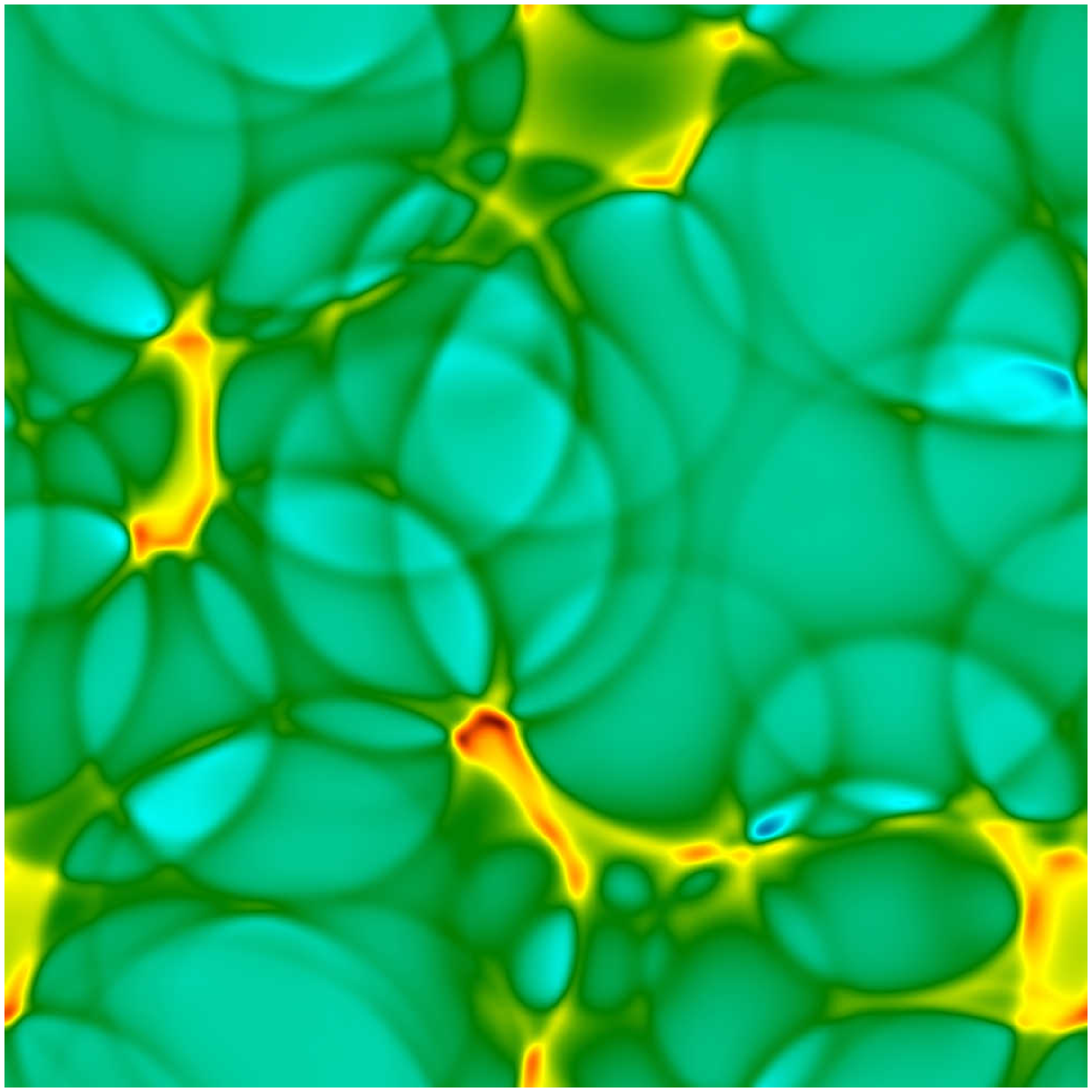}\hspace{5px}
\includegraphics[scale=0.1]{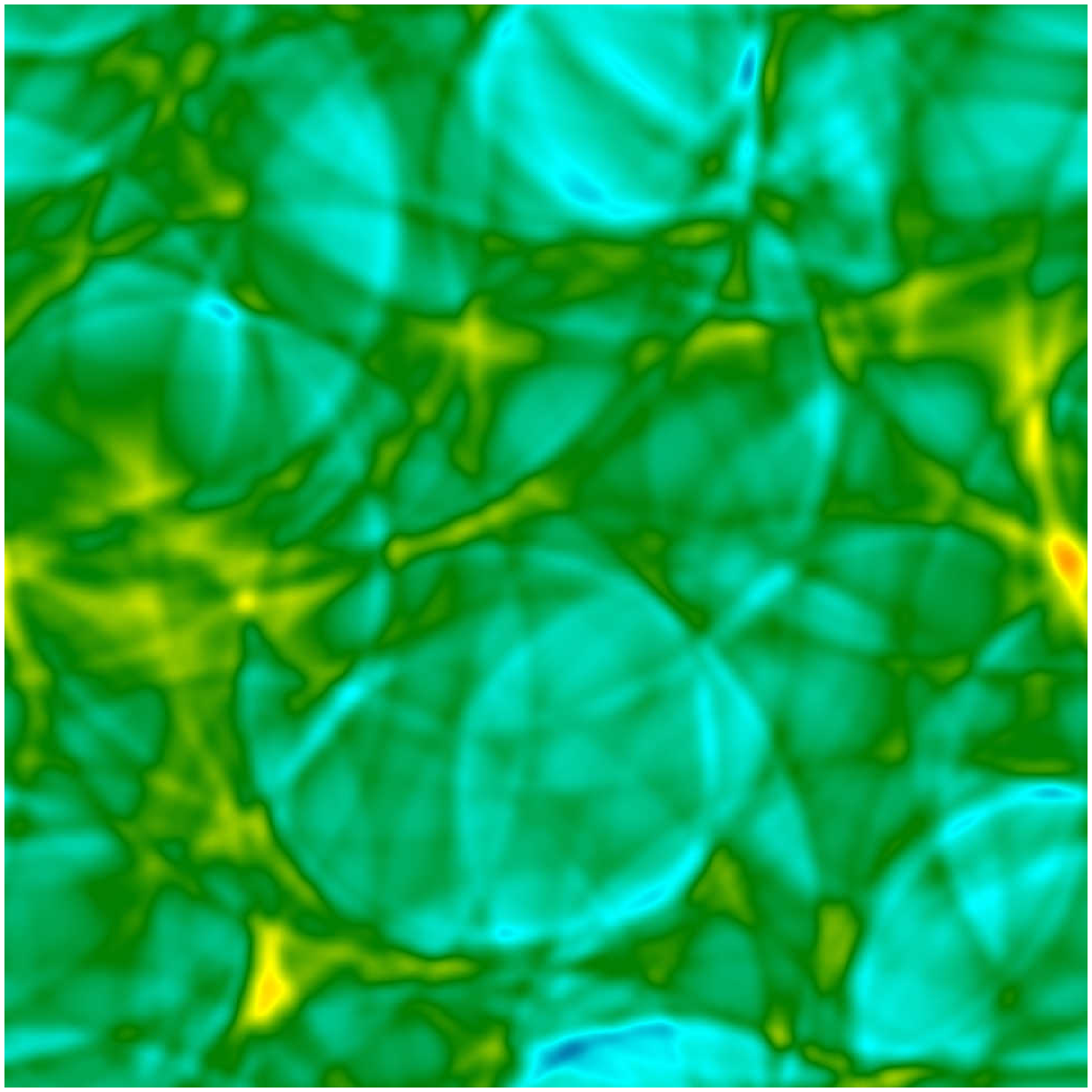}
\caption{\label{fig:slices} Slices of fluid energy density $E/\Tc^4$ at $t=400\, T_\mathrm{c}^{-1}$, $t=800 \,T_\mathrm{c}^{-1}$ and $t=1200\, T_\mathrm{c}^{-1}$ respectively, for the $\eta=0.2$ simulation. The slices
correspond roughly to the end of the nucleation phase, the end of the
initial coalescence phase and the end of the simulation.}
\end{centering}
\end{figure}

The principal observable of interest to us is the power spectrum of
gravitational radiation resulting from bubble collisions.  One
approach is to project $T_{ij}$ at every timestep and then making use
of the Green's function to compute the final power
spectrum~\cite{Khlebnikov:1997di,Easther:2006gt}; this is quite costly
in computer time. Instead, we use the procedure detailed in
Ref.~\cite{GarciaBellido:2007af}. We evolve the equation of motion for
an auxiliary tensor $u_{ij}$,
\begin{equation}
\ddot{u}_{ij} - \nabla^2 u_{ij} = 16 \pi G (\Tfield_{ij} + \Tfluid_{ij}),
\end{equation}
where $\Tfield_{ij} = \partial_i \phi \partial_j \phi$ and
$\Tfluid_{ij} = \relgamma^2 (\epsilon + p)V_i V_j.$ The physical
metric perturbations are recovered in momentum space by $h_{ij}
(\mathbf{k}) = \lambda_{ij,lm} (\hat{\mathbf{k}}) u_{lm}
(t,\mathbf{k})$, where $\lambda_{ij,lm} (\hat{\mathbf{k}})$ is the
projector onto transverse, traceless symmetric rank 2 tensors.  We are
most interested in the metric perturbations sourced by the fluid, as
the fluid shear stresses generally dominate over those of the scalar
field, although it will be instructive to also consider both sources
together.

Having obtained the metric perturbations, the power spectrum per
logarithmic frequency interval is
\begin{multline}
\frac{d\rho_\text{GW}(k)}{d\ln k}
 = \frac{1}{32\pi G L^3} \frac{k^3}{(2\pi)^3} \int d\Omega \, \left| \dot{h}_{lm} (t,\mathbf{k})\right|^2.
\end{multline}
We simulate the system on a cubic lattice of $N^3 = 1024^3$ points,
neglecting cosmic expansion which is slow compared with the transition
rate. The fluid is implemented as a three dimensional relativistic
fluid~\cite{WilsonMatthews}, with donor cell advection. The scalar and
tensor fields are evolved using a leapfrog algorithm with a minimal
stencil for the spatial Laplacian.  Principally we used lattice
spacing $\delta x = 1\, \Tc^{-1}$ and time step $\delta t = 0.1\,
\Tc^{-1}$, where $\Tc$ is the critical temperature for the phase
transition.  We have checked the lattice spacing dependence by
carrying out single bubble self-collision simulations for $L^3 =
256^3\, \Tc^{-3}$ at $\delta x = 0.5\, \Tc^{-1}$, for which the value
of $\rho_\mathrm{GW}$ at $t=2000\Tc^{-1}$ increased by $10\%$, while
the final total fluid kinetic energy increased by $7\%$.  Simulating
with $\delta t = 0.2\, \Tc^{-1}$ resulted in changes of $0.3\%$ and
$0.2\%$ to $\rho_\mathrm{GW}$ and the kinetic energy respectively.

Starting from a system completely in the symmetric phase, we model the
phase transition by nucleating new bubbles according to the rate per
unit volume $P = P_0 \exp (\beta(t-t_0))$.  From this distribution we
generate a set of nucleation times and locations (in a suitable
untouched region of the box) at each of which we insert a static
bubble with a gaussian profile for the scalar field.  The bubble
expands and quickly approaches an invariant scaling profile
\cite{KurkiSuonio:1995vy}.

We first studied a system with $\pdof = 34.25$, $\quadPar= 1/18$,
$\cubPar=\sqrt{10}/72$, $T_0 = \Tc/\sqrt{2}$ and $\lambda = 10/648$;
this allows comparison with previous $(1+1)$ and spherical studies of
a coupled field-fluid system where the same parameter choices were
used \cite{KurkiSuonio:1995vy}.  The transition in this case is
relatively weak: in terms of $\strengthPar{T}$, the ratio between the
latent heat and the total thermal energy, we have $\strengthPar{T_N} =
0.012 $ at the nucleation temperature $\TN = 0.86 \, \Tc$.  We also
performed simulations with $\quadPar= 2/18$ and $\lambda = 5/648$, for
which $\strengthPar{T_N} = 0.10 $ at the nucleation temperature $\TN =
0.8 \, \Tc$, which we refer to as an intermediate strength transition.
We note that $\strengthPar{T_N} \sim 10^{-2}$ is generic for a first
order electroweak transition, while $\strengthPar{T_N} \sim 10^{-1}$
would imply some tuning \cite{Huber:2007vva}.

For the nucleation process, we took $\beta=0.0125 \, \Tc$, $P_0=0.01$
and $t_0 = t_\text{end} = 2000 \, \Tc^{-1}$.  The simulation volume
allowed the nucleation of 100-300 bubbles, so that the mean spacing
between bubbles was of order $100\, \Tc^{-1}$.  The wall velocity is
captured correctly, but the fluid velocity did not quite reach the
scaling profile before colliding.  Typically, the peak velocity prior
to collision is 20-30\% below the scaling value for the deflagrations.

For the weak transition we chose $\eta = 0.1$, $0.2$, $0.4$ and
$0.6$. The first gives a detonation with wall speed $\vw \simeq 0.71$,
and the others weak deflagrations with $\vw \simeq 0.44$, $0.24$, and
$0.15$ respectively.  The shock profiles are found in Figs.\ 2 and 3
of Ref.~\cite{KurkiSuonio:1995vy}; slices of the total energy density
for one of our simulations are shown in Fig.~\ref{fig:slices}. The
intermediate transition was simulated at $\eta = 0.4$, for which the
wall speed is $\vw \simeq 0.44$, very close to the weak transition
with $\eta = 0.2$.

\begin{figure}[t]
\begin{centering}
\includegraphics[trim=2.5cm 0.2cm 0.75cm 0.1cm,clip=true,scale=0.33,angle=270]{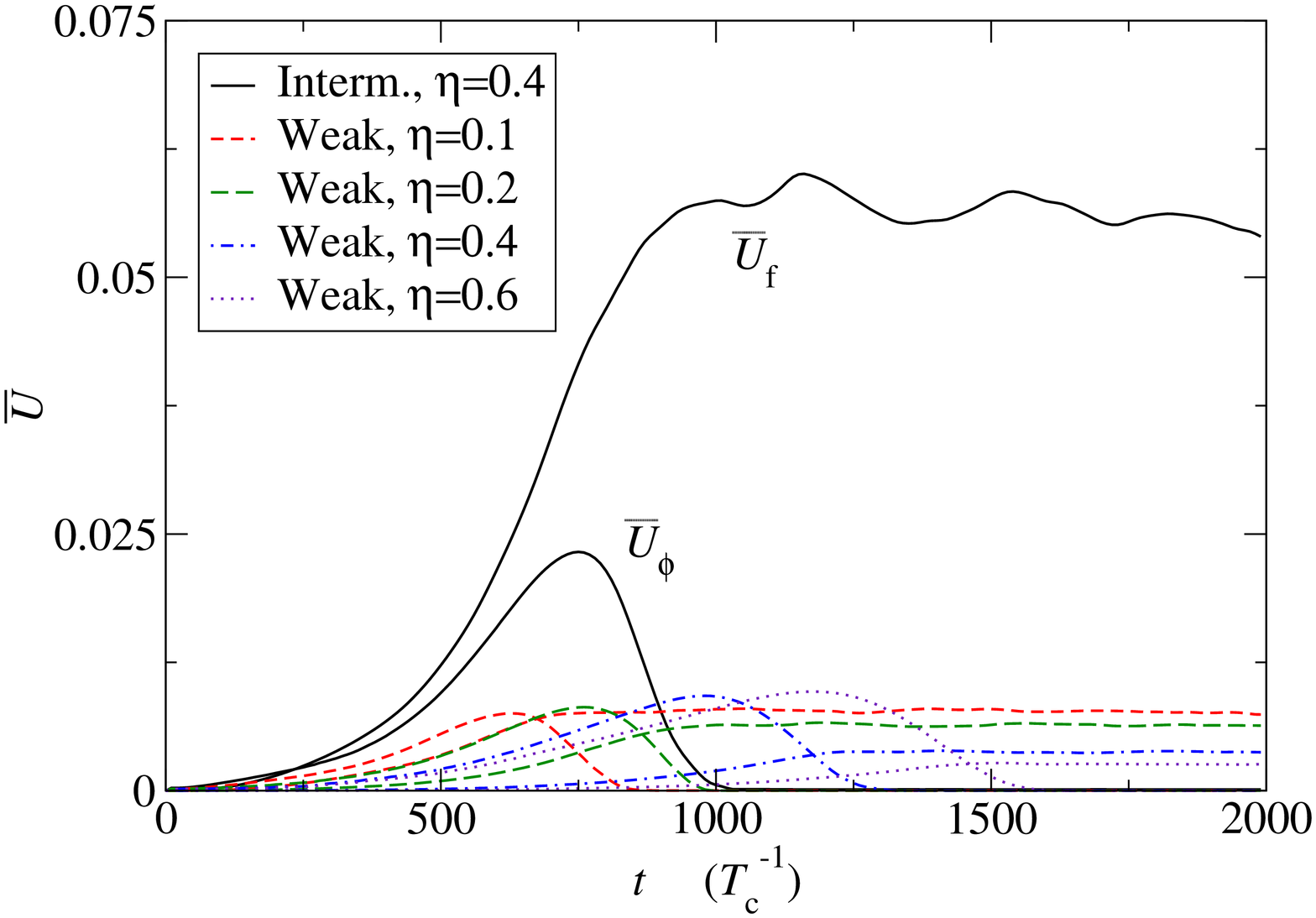}\\
\includegraphics[trim=2.5cm 0.2cm 0.75cm 0.1cm,clip=true,scale=0.33,angle=270]{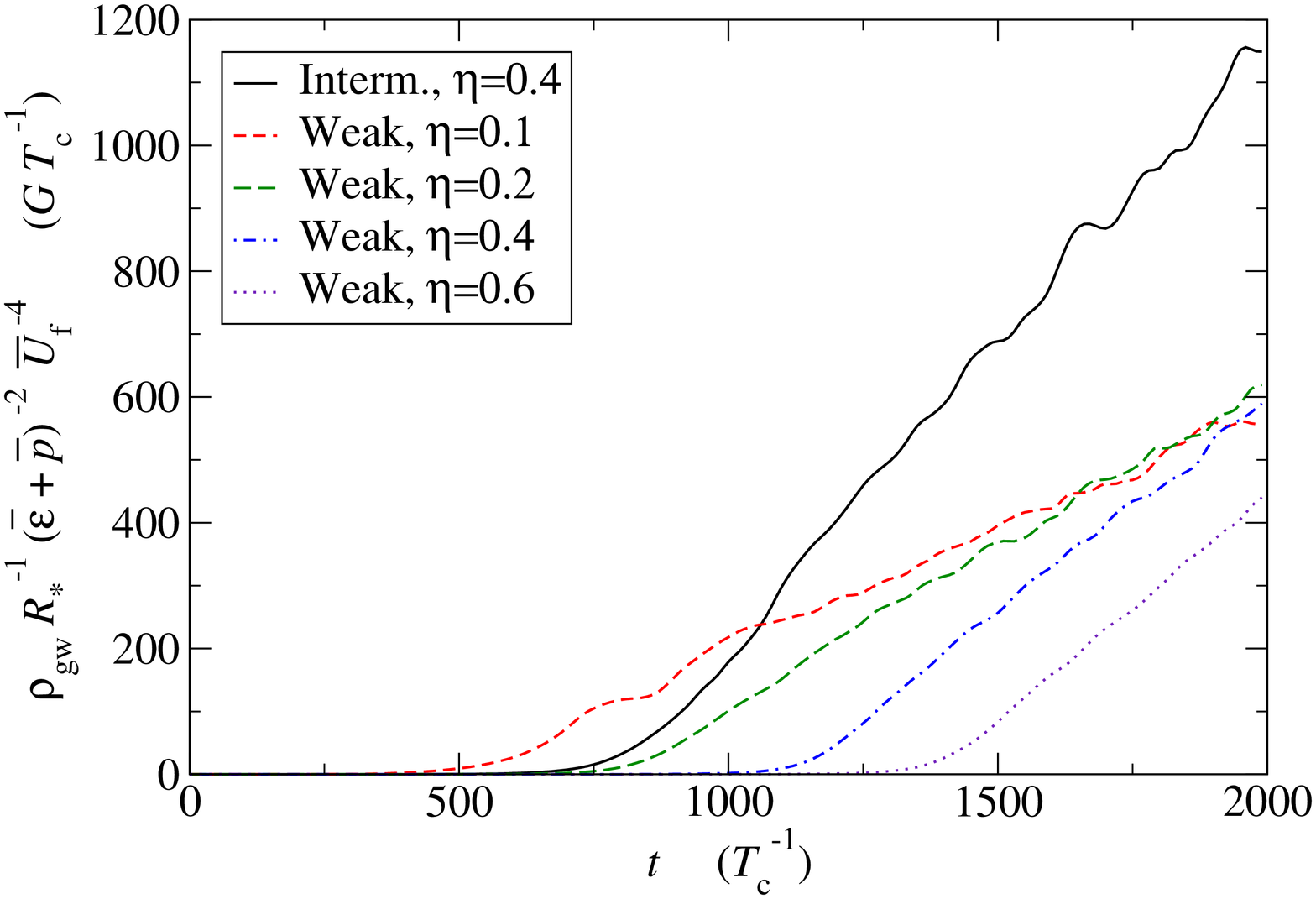}\\
\caption{\label{fig:timevol} Top: time series of $\fieldV$ and $\fluidV$  (\ref{e:Vdefs}), showing the progress of the phase transition; the curves for $\fieldV$ and $\fluidV$ are individually identified for the `intermediate' case. Bottom: time series of $\rho_\text{GW} R_*^{-1} [(\bar \ep+ \bar p)^{-2}\fluidV^{-4}]_{t_\text{end}}$, showing the evolution of the gravitational wave energy density relative to an estimate of the square of the final fluid shear stresses.}
\end{centering}
\end{figure}

Fig.\ ~\ref{fig:timevol} (top) shows the time evolution of two quantities $\fieldV$ and $\fluidV$, defined so that 
\ben
\label{e:Vdefs}
(\bar\ep+\bar p) \fieldV^2 =  \frac{1}{V}\int d^3x\Tfield_{ii} \quad \text{and} \quad (\bar\ep+\bar p)\fluidV^2 =  \frac{1}{V}\int d^3x\Tfluid_{ii}
\een
where $\bar\ep$ and $\bar p$ are the time-dependent, volume-averaged rest-frame energy density and pressure respectively.

The squares of these quantities give an estimate of the size of the
shear stresses of the field and the fluid relative to the background
fluid enthalpy density, while $\fluidV$ tends to the r.m.s.\ fluid
velocity for $\fluidV \ll 1$.  We see that $\fieldV$ grows and decays
with the total surface area of the bubbles of the new phase, while the
mean fluid velocity grows with the volume of the bubbles, and then
stays constant once the bubbles have merged. We have no explicit
viscosity, and the slight decreasing trend in $\fluidV$, visible for
the intermediate transition, arises from the well-known numerical
viscosity of donor-cell advection, $\nu_\text{num} \simeq \fluidV
\delta x$.

Fig.\ ~\ref{fig:timevol} (bottom) shows the GW energy density scaled
by the final value of $(\bar\ep+\bar p)^2\fluidV^4$ and the average
bubble size at collision $R_*= L/\Nb^{1/3}$, where $\Nb$ is the
number of bubbles in the simulation volume.  The scaling enables
comparison to a model discussed around Eq.~(\ref{e:OmgwEqn}), which
predicts a linear growth in $\rho_\text{GW}$ at late times, sourced by
persistent perturbations in the fluid.  The GW energy density rises
linearly after the bubbles have fully merged with similar slopes,
which supports the model.  Note that the GWs from detonations ($\eta =
0.1$) behave similarly to those from deflagrations.

\begin{figure}[t]
\begin{centering}
\includegraphics[trim=2.5cm 0.2cm 1.1cm 0.1cm,clip=true,scale=0.33,angle=270]{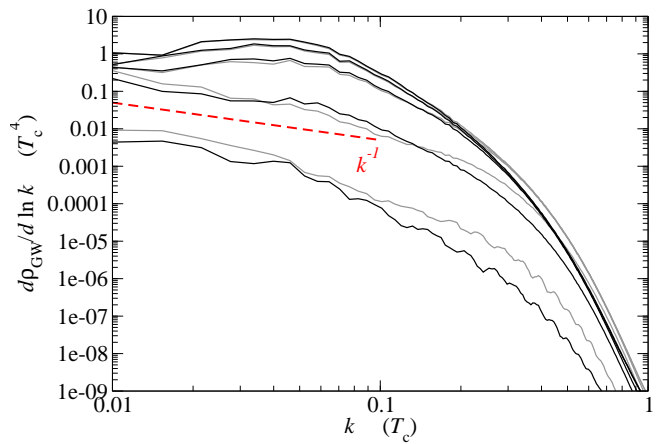}
\caption{\label{fig:GWstack-int} 
Gravitational wave power spectra during the phase transition, for the intermediate strength transition, from fluid only (black) and both fluid and field (grey). From bottom to top, the times are $t = 600$, $800$, $1000$, $1200$ and $1400 \, T_\mathrm{c}^{-1}$. The red dashed line indicates the expected $k^{-1}$ behaviour.}
\end{centering}
\end{figure}

In Fig.\ ~\ref{fig:GWstack-int} we show the time development of the GW
power spectrum as the intermediate strength phase transition
proceeds. We see that strong growth happens between
$t=600\,T_\mathrm{c}^{-1}$ and $t=1000\,T_\mathrm{c}^{-1}$ as the
bubbles merge (see Fig.\ \ref{fig:timevol}). For $t\lesssim
1000\,T_\mathrm{c}^{-1}$ there is evidence of the expected $k^{-1}$
power spectrum, but it becomes less clear as the GW power continues to
grow, sourced by the persistent fluid perturbations.  At the shortest
length scales, we see a $v_\mathrm{w}$-dependent exponential fall-off.

\begin{figure}[t]
\begin{centering}
\includegraphics[trim=2.5cm 0.2cm 1.1cm 0.1cm,clip=true,scale=0.33,angle=270]{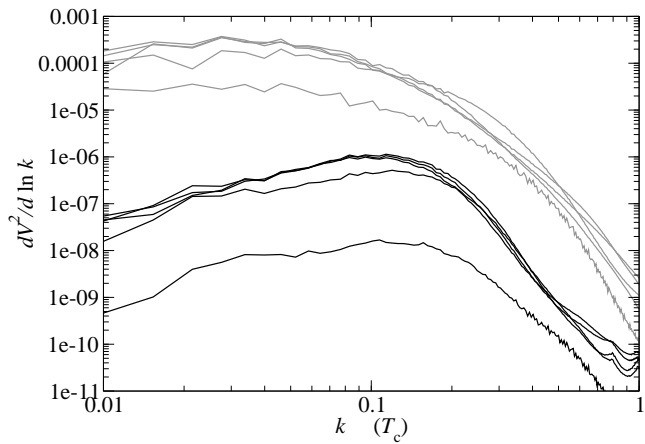}
\caption{\label{fig:fluidcpts} Fluid velocity power spectra for the intermediate strength transition, separated into longitudinal (compressional) and transverse (rotational) components; shown in grey and black respectively. Times shown are the same as Fig.\ \ref{fig:GWstack-int}.  }
\end{centering}
\end{figure}

To establish the nature of these fluid perturbations, we show in
Fig.\ \ref{fig:fluidcpts} the time development of the longitudinal
(compressional) and transverse (rotational) components of the fluid
velocity power spectrum.  At all times, it is clear that most of the
fluid velocity is longitudinal, indicating that the perturbations are
mostly compression waves.  Turbulence generally develops at high
Reynolds number Re in the transverse components, characterised by a
power-law behaviour of the power spectrum. Given the bubble separation
scale $\Rbc$, we can estimate the value of Re, due entirely to the
numerical viscosity, as $\text{Re}_\text{num} = \fluidV
\Rbc/\nu_\text{num} \sim 10^2$.  There is no firm evidence of a power
law at high $k$, but it is unclear whether Re is large enough for
turbulence to develop here.

We can now form a clearer picture of the fluid perturbations and how
the GWs are generated. Firstly, we note that the fluid perturbations
are initially the form of a compression wave surrounding the growing
bubble. The energy in this wave is proportional to the volume of the
bubble $\Rb^3$, and quickly outstrips the energy in the scalar field,
which grows only as $\Rb^2$. The energy in the compression waves
remains constant after the bubbles have merged.  This is due to
linearity and conservation of energy: as the fluid velocities are
generally small, there is little transfer to the transverse
components.

The bubble collision generates gravitational waves, as predicted by
the envelope approximation, and there is some evidence for the
characteristic $k^{-1}$ spectrum between $\Rbc$ and the high-frequency
cut-off.  The generation of GWs continues long after the merger is
completed and the scalar field has relaxed to its new equilibrium
value. The GWs are sourced by the compression waves in the fluid. This
source of gravitational radiation from a phase transition -- sound --
has not been appreciated before (except in
Ref.~\cite{1986MNRAS.218..629H}).

The resulting density of the gravitational waves is given from the
unequal time correlator of the shear stress tensor $\Pi^2(k,t_1,t_2)$
by \cite{Caprini:2009fx,Figueroa:2012kw}
\begin{equation}
\frac{d\rho_\text{GW}(k)}{d\ln k} = \frac{2G k^3}{\pi} \int^t dt_1 dt_2 \cos[k(t_1 - t_2)] \Pi^2(k,t_1,t_2).
\end{equation}
We model the source as turning on at the nucleation time $\tN$ with a
lifetime $\Ts$ (discussed below), and being a function of $t_1-t_2$
between those times, as is reasonable for stochastic sound waves. We
suppose the correlator is peaked at $t_1- t_2=0$ with width
$\tCohPar/k$, where $\tCohPar$ is a dimensionless parameter.  This
resembles the ``top-hat'' correlator model of
Ref.\ \cite{Caprini:2009fx}, except that the source acts for much
longer than the duration of the transition $\be^{-1}$.  We estimate
the amplitude of the source as $\left[(\bar\ep+\bar
  p)\fluidV^2\right]^2$, and its length scale as $\Rbc$.  Hence, for
$\tN < (t_1,t_2) < \Ts$,
\ben
\Pi^2(k,t_1,t_2) \simeq [(\bar\ep+\bar p)\fluidV^2]^2\Rbc^3\tilde\Pi^2(k\Rbc, z/\tCohPar),
\een
where $z = k(t_1 - t_2)$ and $\tilde\Pi^2$ is dimensionless.  The
density parameter $\Om_\text{GW} = \rho_\mathrm{GW}/{\bar \epsilon}$
is then
\begin{equation}
\Om_\text{GW} \simeq \frac{3\bar \Pi^2}{4\pi^2}(\Hc\Ts)(\Hc\Rbc)(1+w)^2\fluidV^4,
\label{e:OmgwEqn}
\end{equation}
where $\Hc$ is the Hubble parameter at the transition, $w =
\bar{p}/\bar{\ep} \simeq 1/3$, and
\ben
\label{e:PiBarDef}
\bar \Pi^2 = \int d\ln k\, (k\Rbc)^2 \! \int dz \cos(z) \tilde\Pi^2(k\Rbc,z/\tCohPar).
\een
In Eq.~(\ref{e:OmgwEqn}) we see the origin of the $\Rbc$ factor in the GW
density, which must be present for dimensional reasons. The slope of
the curves in Fig.\ (\ref{fig:timevol}, bottom) is $2\bar\Pi^2/\pi$,
which we see takes the natural value O(1), and is weakly dependent on
the transition parameters.

The envelope approximation gives \cite{Huber:2008hg}
\begin{equation}
\Om_\text{GW} \simeq \frac{0.11 \vw^3}{0.42+\vw^2} \left( \frac{\Hc}{\be}\right)^2 \frac{\ka^2 \al_T^2}{(\al_T+1)^2} 
\end{equation}
where $\kappa$ is the efficiency with which latent heat is converted
to kinetic energy.  Comparing to (\ref{e:OmgwEqn}) and noting that
$\fluidV^4 \sim \kappa^2\al_T^2$, $\Rbc \sim \vw/\be$, we see that
sound waves are parametrically larger by the factor $\Ts/\Rbc \vw$.

An upper bound on $\Ts$ is the Hubble time, as the shear stresses
decay faster than the background energy density. The shear stresses
also decay due to the viscosity $\etaS$, which can be estimated as
$\etaS \sim T^3/e^4\ln(1/e)$, where $e$ is the electromagnetic gauge
coupling~\cite{Arnold:2000dr}. The viscous damping time of sound waves
with characteristic wavelength $\Rbc$ is therefore $\Tvisc \simeq
\Rbc^2\bar\ep/\etaS \sim e^4\ln(1/e)\Rbc^2\Tc$.  Hence sound waves
from smaller bubbles are damped by viscosity, but live long enough to
be the most important source of gravitational waves for bubbles
provided
\ben
\Rbc\Hc \gg \vw(\sqrt{a}\Tc/\mpl e^4) \sim 10^{-11} \vw (\Tc/100\; \text{GeV}).
\een
This is generally satisfied except for weak transitions at very high
temperatures, and we conclude that for most transitions the fluid
damping time is the Hubble time.

We point out that we have studied systems with non-relativistic and
linear fluid velocities, without explicit viscosity. These choices are
representative of a typical first order electroweak phase transition,
but it would also be interesting to study strong transitions with
relativistic fluid velocities, explore the effect of dissipation, and
look for turbulent regimes. Parameter choices recently identified as
having unstable bubble walls~\cite{Megevand:2013yua} also merit
investigation. We have not studied the case where the walls run away,
although here we expect that the fluid is unimportant and the envelope
approximation applies.

In the cases that we do study, we find the velocity perturbations are
principally acoustic waves, and that the resulting gravitational
radiation density is parametrically larger than given in the envelope
approximation by the ratio of the fluid damping time $\Ts$ to the
duration of the phase transition $\be^{-1}$. We conclude that, for a
wide range of first order phase transitions of interest, the main
source of the gravitational wave background is the sound they make.

\begin{acknowledgments}
Our simulations made use of facilities at the Finnish Centre for
Scientific Computing CSC, and the COSMOS Consortium supercomputer
(within the DiRAC Facility jointly funded by STFC and the Large
Facilities Capital Fund of BIS). KR acknowledges support from the
Academy of Finland project 1134018; MH and SH from the Science and
Technology Facilities Council (grant number ST/J000477/1).
\end{acknowledgments}

\bibliography{fluid}

\end{document}